\documentclass[twocolumn,preprintnumbers,aps,prd,superscriptaddress,nofootinbib,tightenlines]{revtex4-1}

\usepackage{amsmath, amsfonts, amsthm, amssymb, graphicx}
\usepackage{xcolor}
\usepackage[utf8]{inputenc}
\usepackage{hyperref} 
\def\be{\begin{equation}}
\def\ee{\end{equation}}
\def\ba{\begin{eqnarray}}
\def\ea{\end{eqnarray}}

\def\bq{\begin{quote}}
\def\eq{\end{quote}}

\usepackage{colortbl}
\definecolor{summersky}{cmyk}{0.71,0.33,0,0.5}
\definecolor{flamingo}{cmyk}{0,0.51,0.71,0.5}
\definecolor{rp}{cmyk}{0.2, 1, 0.6, 0}
\definecolor{pacificblue}{cmyk}{0.95,0.3,0, 0.5}
\definecolor{gray60}{cmyk}{0.4,0.4,0,0.8}

\hypersetup{
    pdftoolbar=true,        
    pdfmenubar=true,        
    pdffitwindow=true,     
    pdfstartview={FitH},    
    pdfnewwindow=true,      
    colorlinks=true,       
    linkcolor=pacificblue,          
    citecolor=flamingo,        
    filecolor=magenta,      
    urlcolor=pacificblue           
}

\def\n{\nu}

\newcommand{\at}{-\Delta}

 at 10truept

\newcommand{\beq}{\begin{equation}}
\newcommand{\eeq}{\end{equation}}
\newcommand{\bea}{\begin{eqnarray}}
\newcommand{\eea}{\end{eqnarray}}
\newcommand{\beqa}{\begin{eqnarray}}
\newcommand{\eeqa}{\end{eqnarray}}

\newcommand{\then}{\quad \Rightarrow\quad}

\begin{document}
\raggedbottom

\title{Symmetric Superfluids}

\author{Enrico Pajer} 
\email{enrico.pajer@gmail.com}
\affiliation{DAMTP, Centre for Mathematical Sciences, University of Cambridge, CB3 0WA, United Kingdom}

\author{David Stefanyszyn} 
\email{d.stefanyszyn@rug.nl}
\affiliation{Van Swinderen Institute for Particle Physics and Gravity, University of Groningen, \\ Nijenborgh 4, 9747 AG Groningen, The Netherlands}

\date{\today}

\begin{abstract}
We present a complete classification of symmetric superfluids, namely shift-symmetric and Poincar\'e invariant scalar field theories that have an enlarged set of classically conserved currents at leading order in derivatives. These theories arise in the decoupling limit of the effective field theory of shift-symmetric, single-clock cosmologies and our results pick out all models with couplings fixed by additional symmetry. Remarkably, in $  D\geq 2 $ spacetime dimensions there are only two possibilities: the Dirac-Born-Infeld theory and Scaling Superfluids with Lagrangian $  (-\partial_{\mu}\phi\partial^{\mu}\phi)^{\alpha}$, for some real $ \alpha$. The scaling symmetry present for any $  \alpha $ is further enhanced to the full conformal group only for $  \alpha=D/2 $, and to infinitely many additional generators for the cuscuton, namely $  \alpha=1/2 $. We discuss the stability of Scaling Superfluids and point out that all coupling constants are determined by the speed of sound. 

\end{abstract}

\maketitle

\section{Introduction}

Spacetime symmetries\footnote{By ``spacetime symmetries'' we denote symmetries that do not commute with the Poincar\'e group.} beyond the Poincar\'e group are highly constrained. Under some reasonable assumptions, Coleman and Mandula \cite{Coleman:1967ad} proved that the symmetry group of scattering amplitudes must be a direct product of some internal group and the Poincar\'e group. Two prominent ways are known to avoid this no-go result by invalidating its assumptions. The first is to allow some symmetry generators to obey anti-commutation relations \cite{Haag:1974qh}. This leads to supersymmetry and its infinitely rich structure. The second is to consider theories that do not admit an S-matrix. This is the case for Conformal Field Theories (CFT's), which provide us with crucial landmarks in the space of all Quantum Field Theories. Here, we highlight a third interesting and generic way to avoid the assumptions of the Coleman-Mandula theorem, namely symmetry transformations that do not map one-particle states into one-particle states. This is generally the case for spontaneous symmetry breaking, when symmetry transformations are non-linearly realised. We find this possibility particularly compelling since many physically interesting systems break some Poincar\'e symmetries, e.g.~ Lorentz boosts in cosmology and condensed matter. 

In this paper, we start exploring one of the simplest classes of theories: a single, shift-symmetric scalar field $\phi(x)$, and restrict ourselves to leading order in derivatives i.e.~one derivative per field. This scalar is the Goldstone mode of a spontaneously broken $U(1)$ symmetry (see e.g.~\cite{Runaway} for a top-down discussion). We require the theory to be Poincar\'e invariant but allow for this symmetry to be spontaneously broken. These theories are also known as $  P $-of-$  X $ theories with the Lagrangian an arbitrary function of $X = -\partial_{\mu}\phi \partial^{\mu}\phi$. A particularly interesting class of these theories is Superfluids where the non-linearly realised $U(1)$ symmetry is at finite charge density. This occurs on the Lorentz breaking vacuum $\langle \phi(x) \rangle = \mu t$ and the resulting theory of fluctuations describes the decoupling limit of the effective field theory of shift-symmetric, single-clock cosmologies, corresponding to the subHubble regime.

Superfluids are introduced in Section \ref{sec2} where we classify all possible new symmetries for superfluids that form a consistent algebra with Poincar\'{e} symmetries and the $U(1)$ shift symmetry. Remarkably, we find that in $  D\neq 1 $ spacetime dimensions there are only two possibilities: the Dirac-Born-Infeld action and Scaling Superfluids where the Lagrangian is a monomial in $X$: $P(X) = X^{\alpha}$. As summarized\footnote{In our notation for the semi-direct product $ \rtimes$, the normal subgroup $ N$ is on the left-hand side, $ G=N\rtimes H$}  in Table \ref{tab}, for $  \alpha=D/2 $ (Conformal Superfluid \cite{Herzog:2008he,Esposito:2016ria}) and $  \alpha=1/2 $ (Cuscuton \cite{Afshordi:2006ad,Afshordi:2007yx}) the scaling symmetry is enhanced to the full conformal group and to additional vector and scalar generators (see \eqref{jacobi}, \eqref{Wgen}), respectively. 

\begin{table}[h!]
\begin{center}
\begin{tabular}{ |c|c| } 
\hline
$ P(X)$ & Symmetries\\ 
\hline \hline
 $\sqrt{1-X}$ & $\text{ISO}(D,1)$   \\ 
 \hline
$ X^{\alpha}$ & $U(1) \rtimes(\mathbb{R}^{D}\rtimes(SO(d,1)\times \mathbb{R}))$   \\ 
 \hline
$ X^{D/2}$ & $U(1)\times SO(D,2) $  \\ 
 \hline
 $ X^{1/2}$ & $ \mathbb{R}^{\infty}\rtimes ( U(1) \rtimes(\mathbb{R}^{D}\rtimes(SO(d,1)\times \mathbb{R}))) $  \\ 
 \hline
\end{tabular}
\caption{All interacting, Poincar\'{e}-invariant, shift-symmetric theories with enhanced symmetry in $  D=d+1 $ spacetime dimensions, to leading order in derivatives. See the end of Sec. \ref{Tadpoles} for a discussion of possible tadpoles. \label{tab}}
\end{center}
\end{table}

It is perhaps at first surprising to find scaling but not conformal symmetry as is the case for $\alpha \neq D/2$, but this is compatible with all results in the literature (e.g.~ \cite{Jackiw:2011vz,Dorigoni:2009ra,ScaleVConformal1,Polchinski,Fortin:2012hn,ScaleVConformal3,ScaleVConformal4}) which rely on linearly realized symmetries. Here the scaling symmetry is non-linearly realised on the Superfluid phonon which is the scalar fluctuation around the Lorentz violating vacuum. In addition, the classical scaling symmetry is in general anomalous. Interestingly, Scaling Superfluids are very restricted theories in which all interactions, with a single derivative per field, are fixed by the speed of sound.

Importantly, in our analysis we do not assume the existence of a sensible theory of fluctuations around the Poincar\'{e} invariant vacuum $\langle \phi(x) \rangle = \text{const}$ since ultimately we are interested in Lorentz breaking vacua. For this reason, our results extend those reported in a classification of scalar EFT's based on their amplitudes' soft scaling \cite{Adler1,Adler2,ScalarSoft1,ScalarSoft2,ScalarSoft3,ScalarLie1}. We find additional possibilities, namely Scaling (and Conformal) Superfluids, which do not admit a perturbative S-matrix when the Poincar\'{e} symmetries are unbroken. Indeed, they possess vacua that are either Poincar\'e invariant or perturbative, but not both. On the other hand, the classification based on amplitudes (and Lie-algebras \cite{AlgebraicClassification}) is directly applicable to theories with higher derivatives, while for us that requires a whole new analysis, which we postpone to future work. In any case, all symmetries which we find, classified in terms of the number of Lorentz indices of the corresponding generator and the highest power of the spacetime coordinate appearing in the transformation rule, are collected in Table \ref{tab2}.

Two technical caveats of our results are worth mentioning:
\begin{enumerate}
\item We consider only symmetry transformations that are a finite order polynomial in the spacetime coordinate $  x $, as in \eqref{mform}. This excludes conformal transformations in $  D=2 $, which include arbitrary holomorphic coordinate transformations. But our analysis still holds since we are interested in interacting theories and the only Conformal Superfluid in $  D=2 $ is the free theory $  P=X $, as we shall see.
\item We derive all possible symmetric Lagrangians, but we do not derive all possible conserved currents. In particular,  we were not able to rule out new conserved currents for the $  \alpha=1/2 $ Scaling Superfluid beyond those present in Table \ref{tab}. We do, however, in Section \ref{DegreeThreeHigher} show that there are no additional conserved currents for the Conformal Superfluid in $D \geq 3$.
\end{enumerate}

One final apology: our derivation will be completely classical. We will only briefly discuss possible issues that emerge at the quantum level, such as anomalies, in the final discussion. An extension to scattering amplitudes akin to Coleman and Mandula is in principle possible and left for future work.

%

The rest of the paper is organised as follows. In Section \ref{sec2}, we classify all possible Superfluid actions with enhanced symmetry at the classical level. We then study the Scaling and Conformal Superfluids in more detail in Section \ref{sec3}. We study the theories in terms of fluctuations around the Lorentz breaking background, computing the speed of sound and leading order interactions. We also compute higher order corrections which involve more derivatives using the coset construction. We end with a conclusion and discussion of future work.

\begin{table}[t]
\begin{center}
\begin{tabular}{ |c|c|c|c| } 
\hline
 & $ N=0$ & $ N=1$ & $ N=2$ \\ 
\hline 
 $m=0$ & $Q, \mathbb{I}$ &$ \hat D, \, W $ &  \\ 
 \hline
$ m=1$ & $P_{\mu},\,V_{\mu}$ & $A_{\mu}$  & $ K_{\mu}$  \\ 
 \hline
$m=2$ & &$M_{\mu\nu}$ &    \\ 
 \hline
\end{tabular}
\caption{This table summarizes all superfluid symmetries organized in powers or $ x^{N}$ and the number $ m$ of Lorentz indices (not irreps). Commuting with spacetime translations move us leftward along a diagonal. \label{tab2}}
\end{center}
\end{table}


\section{Symmetric Superfluids}\label{sec2}

Although the fundamental laws of Nature appear to be covariant under spacetime translations and Lorentz transformations, the World around us is clearly not Poincar\'e invariant and so some of these symmetries must be broken spontaneously. Indeed, condensed matter systems and cosmology both break Lorentz boosts and should therefore be described in terms of Goldstone modes and matter fields which \textit{non-linearly} realise the full Lorentzian symmetries. The most simple effective field theory (EFT) of this type is that of Superfluid phonons. This EFT consists of a single gapless Goldstone excitation $\pi$ and at low energies the leading order interactions are constructed out of the combination \cite{Son,Superfluid1} 
\begin{align}
\mu \dot{\pi} + \frac{1}{2} \dot{\pi}^{2} - \frac{1}{2}\partial_{i}\pi \partial^{i}\pi\,.
\end{align}
This particular tuning between the three terms is necessary to non-linearly realise boosts. We can re-write the Superfluid action in terms of a new scalar $\phi = \mu t + \pi$. All Poincar\'{e} symmetries are now linearly realised on $\phi$: this makes it clear that the Superfluid is describing a Lorentz breaking phase in a fundamentally relativistic theory. The leading order action for the Superfluid in $D = 1+d$ spacetime dimensions is \cite{Son,Superfluid1} 
\begin{align}
S = \int d^{D}x P(X) \label{Superfluid-action}\,,
\end{align}
where $X = -\partial^{\mu}\phi \partial_{\mu}\phi$. The Superfluid is defined as a state where the spontaneously broken $U(1)$ symmetry has finite charge density which in turn requires $\dot{\phi} \neq 0$. See \cite{FiniteTemperature} for more details. 

This action can be derived from the coset construction \cite{Coset1,Coset2,Coset3}  where the full symmetry group $G$ is $\text{ISO}(1,d) \times U(1)$ and is broken down to a subgroup consisting of spatial rotations, spatial translations and a new form of time translations which corresponds to a linear sum of the original time translations and the $U(1)$ \cite{SuperfluidCoset,Zoology}. Indeed the ground state $\langle \phi(x) \rangle = \mu t$ is invariant under $t \rightarrow t + t_{0}, \phi \rightarrow \phi - \mu t_{0}$. The Goldstone modes corresponding to the spontaneously broken Lorentz boosts can be eliminated from the Goldstone EFT by imposing inverse Higgs constraints \cite{InverseHiggs}, illustrating that for spontaneously broken spacetime symmetries there can be fewer Goldstone modes than broken generators \cite{LowGoldstone}.

For arbitrary $P(X)$ the action \eqref{Superfluid-action} is invariant under Poincar\'{e} symmetries (translations $P_{\mu}$ and Lorentz transformations $M_{\mu\nu}$) and the non-linearly realised $U(1)$ symmetry $Q$. These symmetries respectively act on $\phi$ in the following way\footnote{More generally, we have $ \delta_{G} f(\phi)=f'(\phi)\delta_{G} \phi$ for any function $ f$ and generator $ G$. }
\begin{align}
\delta_{P_{\mu}} \phi = -\partial_{\mu}\phi, \quad \delta_{M_{\mu\nu}} \phi = 2x_{[\mu}\partial_{\nu]}\phi, \quad \delta_{Q}\phi = 1 \,,\label{Superfluid-Symmetries}
\end{align}
and satisfy the non-zero commutation relations
\begin{align}
[M_{\mu\nu},P_{\sigma}] = 2\eta_{\sigma [\mu}P_{\nu]}, \quad [M_{\mu\nu}, M^{\rho\sigma}] = 4\delta_{[\mu}^{[\rho}M_{\nu]}{}^{\sigma]}\,,  \label{Poincare-Algebra}
\end{align}
where we anti-symmetrise with weight one. Since the $U(1)$ generator $Q$ only generates a shift for $\phi$ it commutes with translations. We stress that here and throughout this paper we work with the \textit{active} form of the transformation rules where the spacetime coordinates do not transform. This is opposed to the \textit{passive} form where both the field and coordinates can transform (see e.g.~ \cite{Olver}).

Our primary goal is to find special choices for $P(X)$ that have new symmetries in addition to \eqref{Superfluid-Symmetries} while maintaining interactions\footnote{There are infinitely many symmetries of the free theory \cite{Extended}.}. To keep our results completely general we allow new transformation rules to be a function of the scalar field, its derivatives and the spacetime coordinates. A candidate new transformation for $\phi$, which is generated by a new generator $S_{\mu_{1}, \ldots \mu_{m}}\equiv S_{m}$, takes the form
\begin{align}
\delta_{S_{m}} \phi = f_{\mu_{1}, \ldots \mu_{m}}(x^{\mu}, \phi, \partial_{\mu}\phi, \partial_{\mu}\partial_{\nu}\phi, \ldots) \label{New-Symmetry}\,,
\end{align}
where $S_{\mu_{1}, \ldots \mu_{m}}$ is a covariant tensor under Lorentz transformations. 

Any new symmetries have to form a consistent algebra with $\text{ISO}(1,d) \times \text{U}(1)$ otherwise we would not find invariants which are subsets of \eqref{Superfluid-action}. Following \cite{Haag:1974qh}, we assume the transformation rule is perturbative around $x^{\mu} = 0$ and we define $  \mathcal{G}(N,m) $ to be the set of symmetries $  S^{N}_{m} $ with $ m$ Lorentz indices and of order $ x^{N}$, i.e.~of the explicit form
\begin{align}\label{mform}
\delta_{S^{N}_{m}} \phi \equiv \sum_{n=0}^{N} x^{\mu_{1}}\dots x^{\mu_{n}}  f_{\mu_{1} \dots \mu_{n}\mu_{n+1}\dots \mu_{n+m}} 
\end{align}
for some finite integer $ N $ and $ f$ a local function of $ \phi$ and its derivatives that is non-vanishing for $  n=N $. A few comments are in order. First, $  S^{N}_{m} $ in general form reducible representations of the Lorentz group. One could in principle further organize them into irreducible representations (``spin'' in $  D $ dimensions) but we don't find this particularly useful at this stage and therefore we prefer to keep a more streamlined notation. Second, the $  S^{N}_{m} $ are not necessarily independent from $  S^{N}_{m'} $ with $  m'<m $. For example, one could have $  S^{N}_{m}=\eta_{\mu\nu}S^{N}_{m-2} $, which of course does not generate any new conserved current.

Symmetries that are in $  \mathcal{G}(N,m) $ but not in $  \mathcal{G}(N-1,m) $ are called \textit{symmetries of degree} $  N $. Let's consider the commutator between translations and $S^{N}_{m} \in \mathcal{G}(N,m)$, which at the level of transformation rules reads
\begin{align}
[\delta_{P_{\mu}},\delta_{S^{N}_{m}}] \phi &= \delta_{P_{\mu}}(\delta_{S^{N}_{m}} \phi) + \delta_{S^{N}_{m}}(\partial_{\mu}\phi) \nonumber\\
&=\sum_{n=0}^{N}f_{\mu_{1} \dots \mu_{n}\mu_{n+1}\dots \mu_{n+m}}  \partial_{\mu} \left( x^{\mu_{1}}\dots x^{\mu_{n}} \right)  \nonumber\\
&\equiv \delta_{S_{m+1}^{N-1}}\phi,\label{back}
\end{align}
where the partial derivative acts only on the explicit spacetime dependence, i.e.~at constant $  \phi $. Notice that the coefficient of $  x^{N-1} $ cannot vanish by the definition of $  S^{N}_{m} $. This tells us that 
\begin{align}
[P_{\mu},S_{m}^{N}]=S_{m+1}^{N-1}\in \mathcal{G}(N-1,m+1)\,.
\end{align}
If $ S_{m}^{N}$ is to be a symmetry of the action, there must also be a symmetry $ S_{m+1}^{N-1}$, otherwise the algebra would not close. We therefore find it convenient to classify new symmetries in terms of their explicit $x^{\mu}$ dependence and the number of un-contracted Lorentz indices, similar to what was done in \cite{VectorLie1}. We now study symmetries in increasing order of degree $  N $. In what follows we work in $  D\neq 1 $ spacetime dimensions. The special case of $  D=1 $, corresponding to quantum mechanics, is discussed separately in Subsection \ref{ssec:QM}.


\subsection{Degree zero}

First consider the case $N=0$, namely transformations that commute with $  P^{\mu} $
\begin{align}
\left[  P_{\mu},S^{0}_{m}\right]=0\,.
\end{align} 
This includes translations themselves and the scalar's shift symmetry. These transformations form a subalgebra since by Jacobi identities
\begin{align}
\left[  P_{\mu},\left[  S^{0}_{m},S^{0}_{m'}\right]\right]=0\,. 
\end{align}
By demanding invariance of $P(X)$ we find that the only other possibility is an infinite number of vector generators $S_{1}^{0}=V_{\mu}(f)$, acting as
\begin{align} 
\delta_{V_{\mu}} \phi = f(\phi) \partial_{\mu}\phi\,, \label{SquareRootSymmetry}
\end{align}
where $f(\phi)$ is an arbitrary function of $\phi$. The new commutation relations, in addition to $\eqref{Superfluid-Symmetries}$, are
\begin{align}\label{jacobi}
[Q, V_{\mu}(f)] &=  V_{\mu}(f')\,, &  [J_{\mu\nu},V_{\sigma}] &= 2\eta_{\sigma [\mu}V_{\nu]}\,, \\ \nonumber
  [V_{\mu}(f),V_{\nu}(\tilde f)]&=0 \,, &   [P_{\mu},V_{\nu}]&=0 \,, 
\end{align}
where $  f'=\partial_{\phi}f $.  The invariant action to leading order in derivatives is
\begin{align}
S = \int d^{D}x \sqrt{X}\,,  \label{SquareRoot}
\end{align}
which is simultaneously invariant under the infinite set of transformations generated by $  V_{\mu}(f) $ for any $  f(\phi) $. Under the transformations in \eqref{SquareRootSymmetry}, the Lagrangian changes by a total derivative
\begin{align}
\delta_{b}\mathcal{L}=\partial_{\mu}\left(  b^{\mu}f\sqrt{X}\right)\,,
\end{align}
so the infinitely many conserved currents are
\be
\left( J_{\nu}(f) \right)^{\mu}=-\frac{f}{\sqrt{X}}\left[  \partial^{\mu}\phi \partial_{\nu}\phi+\delta^{\mu}_{\nu}X\right]\,.
\ee
These currents are independent for different choices of $ f$ and non-vanishing. This theory has actually additional symmetries of higher degree (see \eqref{additional}), which we discuss in the next subsection. Surprisingly, this theory nevertheless admits infinitely many solutions. To see this, notice that $ \phi=c_{\mu}x^{\mu}$ is trivially a solution of the equations of motion
\begin{align}\label{eomsqrt}
X\Box \phi =-\partial^{\mu}\phi \partial^{\nu}\phi \partial_{\mu}\partial_{\nu}\phi\,,
\end{align} 
for any constant four-vector $c^{\mu} $. Then we can apply a symmetry transformation to generate a new solution 
\begin{align}
\phi(x)=h(c_{\mu}x^{\mu})
\end{align}
for any function $h $. This theory is non-perturbative around the Poincar\'{e} invariant vacuum $\langle \phi(x) \rangle = \text{const}$ but it is also problematic around the Lorentz breaking vacuum\footnote{Indeed, by expanding \eqref{SquareRoot} around this solution we see that that the symmetry forces the coefficient of $\dot{\pi}^{2}$ to vanish, see also \cite{Tasinato}.} $ \langle \phi(x) \rangle  = \mu t $. On the other hand, it admits a perturbative regime around more general background solutions of the form $  \phi(x)  =c_{\mu}x^{\mu} $ for non-vanishing $ c_{i}$.

The careful reader might have noticed that we did not put an absolute value around $ X$ in the argument of the square root in \eqref{SquareRoot}. Indeed, at the classical level, which is relevant for our discussion, the variational principle based on the action in \eqref{SquareRoot} can still be used to derive the classical equations of motion as we have done, even for solutions with $ X<0$, for which the action becomes imaginary. At the quantum level instead, an imaginary action and Hamiltonian would generically lead to violations of unitarity, and it would be safer to work with $ \sqrt{|X|}$ instead. Nevertheless, as long as we are interested in quantizing the theory around a ``timelike'' background with $ \bar X>0$ as for example with $ \langle \phi(x) \rangle =\mu t$, we don't expect to see any difference in perturbation theory between $ X$ and $ |X|$. Small fluctuations around $ \bar X \neq 0$ probe the Lagrangian only locally and do not reach $ X=0$. Non-perturbative effects might arise from the absolute value, and it would be nice to investigate if they can be trusted within the validity of the EFT and whether they lead to some interesting (and consistent) phenomenology. This discussion applies to all other models discussed in the  rest of this paper, including the well-studied DBI Lagrangian which we will encounter in the next subsection.

 
\subsection{Degree one}\label{ssec:}

We now move onto transformations $  S^{1} $ with $N=1$, where $ \delta \phi$ is at most linear in $ x$ and takes the schematic form (with implicit Lorentz indices)
\begin{align}
\delta_{S^{1}} \phi = f_{0}(\phi, \partial \phi, \ldots) + f_{1}(\phi, \partial \phi, \ldots) x\,.
\end{align}
We know from \eqref{back} that the commutator between this transformation and a translation is non-zero and must give back a symmetry $  S_{m+1}^{0} $. If this degree-zero symmetry $  S^{0} $ is the one in \eqref{SquareRootSymmetry}, then we will recover the result of the previous section, namely \eqref{SquareRoot}. So, to find something new, we impose that the $  S^{0} $ symmetries derived from $  S^{1} $ are translations and the shift symmetry. However, before doing so let us mention that we found an additional symmetry of \eqref{SquareRoot} which is degree one. Indeed, that action is invariant under the following symmetry generated by a new scalar generator $W$:
\begin{align}\label{additional}
\delta_{W}\phi = g(\phi) + \frac{\partial_{\phi} g(\phi)}{D-1}x^{\mu}\partial_{\mu}\phi
\end{align}
where $g(\phi)$ is an arbitrary function of $\phi$ only. The new non-zero commutators involving $W$ are
\begin{align} \label{Wgen}
[P_{\mu},W] = V_{\mu}, \quad [V_{\mu},W] = V_{\mu}.
\end{align} 
It would be interesting to investigate if there are even more symmetries for this Square Root theory but since here we are primarily interested in new theories, we leave that analysis for future work. Clearly this theory deserves further attention.

Now, to find something new we must have
\begin{align}
f_{1}^{\mu} = a_{1}^{\mu} - a_{2}^{\mu\nu} \partial_{\nu} \phi\,,
\end{align}
where $a_{1}^{\mu}$ and $a_{2}^{\mu\nu}$ are real constants. We also know that the commutator between this $N=1$ transformation and the shift symmetry must give back a symmetry of the theory and therefore $\partial f_{0} / \partial \phi$ must also correspond to a shift or a translation. We therefore have
\begin{align}
\delta_{S^{1}} \phi &= f_{0}+ f_{1} x\nonumber\\
&=\phi + \phi \partial_{\mu} \phi +x_{\mu}  - x_{\mu}\partial_{\nu} \phi \,.
\end{align}
Based solely on Lorentz invariance, the parameters $ a=\{a_{\dots}\}$ of an $N=1$ symmetry transformation can therefore be
\begin{align}
a\cdot \delta_{S^{1}} \phi =\phi \left( a_{3}  + a_{4}^{\mu}  \partial_{\mu} \phi \right)+ (a_{1}^{\mu} - a_{2}^{\mu
\nu} \partial_{\nu} \phi) x_{\mu}\,,
\end{align}
i.e.~a scalar, $  a_{3} $ and the trace $  a_{2}\equiv a_{2}^{\mu\nu}\eta_{\mu\nu} $, a vector, $  a_{4}^{\mu} $ and $a_{1}^{\mu}$, or a traceless, symmetric rank-2 tensor $  \left( a^{\mu\nu}_{2}-\eta^{\mu\nu}a_{2}/ D \right) $. The antisymmetric 2-form reduces to a Lorentz transformation. We considered each case and by demanding invariance of $P(X)$ we found that the parameter must be a scalar or a vector, so $  a_{2}^{\mu
\nu}\equiv a_{2}\eta^{\mu\nu}$.

For $ a_{1}^{\mu}=a_{4}^{\mu}=\left( a^{\mu\nu}_{2}-\eta^{\mu\nu}a_{2} /D \right) =0$, the resulting scalar symmetry is recognized as the generator of scaling transformations and we will denote it by\footnote{The hat is introduced to avoid confusion with the spacetime dimension $ D$.} $\hat{D}$. The associated transformation rule is  
\begin{align}
\delta_{\hat{D}}\phi = -\Delta \phi - x^{\mu}\partial_{\mu}\phi\,, \label{Scaling-Symmetry}
\end{align}
where we have set $a_{2} = 1$ without loss of generality and we defined the scaling dimension $  \Delta $ of $  \phi $ by $  \Delta\equiv-a_{3}$. Invariance of the action then reduces it to a monomial in $X$:
\begin{align}
S = \int d^{D}x X^{\alpha}\,, \label{Scaling-Superfluid} \quad \text{with} \quad \alpha = \frac{D }{2(1+\Delta)} \,,
\end{align}
for $  \Delta\neq -1 $. At this order in derivatives there are no invariant theories with $  \Delta=-1 $ because $  X $ is scale invariant and it not possible to offset the scaling of $  d^{D}x $.

The non-zero commutation relations for this new algebra are \eqref{Poincare-Algebra} plus the new commutators\footnote{Notice that the Jacobi identities are satisfied for this algebra as well as for \eqref{jacobi}.}
\begin{align}
[P_{\mu},\hat{D}] =  P_{\mu}\,, \qquad [Q,\hat{D}] = -\Delta Q.
\end{align}
The first of these commutators tells us that $\hat{D}$ is the generator of dilatations while the second tells us that $Q$ transforms under dilatations with scaling weight $-\Delta$. Although this symmetry acts linearly on $\phi$, it acts non-linearly on the Superfluid phonon $\pi$ and as we shall see shortly, the symmetry fixes the speed of sound and all $\pi$ self-interactions in terms of the single number $\Delta$. We note that different values of $\Delta$ correspond to different algebras and therefore physically different Superfluid theories. We denote a Superfluid theory described by this action as a \textit{Scaling Superfluid} for obvious reasons and will study its properties in more detail in the following section.

Lastly, we have the case when the degree one symmetry is generated by a vector, which we denote as $A_{\mu}$. By definition, one of the four components of  $a_{1}^{\mu}$ must be non-vanishing, otherwise the symmetry would not have degree one. Then by Lorentz invariance the symmetry should exist also for a Lorentz transformed $ a_{1}^{\mu}$, for which all components are in general non-vanishing. We can then rescale $ a_{1}^{\mu}=1$ without loss of generality, such that the transformation rule is 
\begin{align}
\delta_{A_{\mu}} \phi = x_{\mu} + a_{4} \phi \partial_{\mu}\phi. \label{DBIsymmetry}
\end{align}
For $a_{4} = 0$ the transformation reduces to a Galileon transformation \cite{Galileon} and in this case the only $P(X)$ invariant is that of the free theory\footnote{There are of course Galileon invariants, which involve more derivatives than $P(X)$ interactions, and the leading order ones are Wess-Zumino terms \cite{Goon}.}. Conversely, for $a_{4} \neq 0$ we can set $a_{4} = 1$ by a rescaling of the transformation and a field redefinition. Without loss of generality then the invariant action takes the form\footnote{Note again that both at the classical level and at the perturbative quantum level away from $ X=1$ we do not worry about the solutions that make the action imaginary.}
\begin{align}
P = \sqrt{1 -  X} \,.\label{DBI-action}
\end{align}
This is the well known Dirac-Born-Infeld (DBI) action which non-linearly realises the $(D+1)$-dimensional Poincar\'{e} group. The $U(1)$ symmetry $Q$ then corresponds to translations in the extra dimension, i.e.~$ Q=P_{D+1}$, while the new vector generator $A_{\mu}$ corresponds to Lorentz transformations involving the extra dimension, i.e.~$ A_{\mu}=M_{\mu(D+1)}$. The non-zero commutators are the generalisation of \eqref{Poincare-Algebra} to one extra \textit{spatial} dimension. This theory requires no further discussion but let us mention that higher derivative corrections were considered in \cite{Reunited} and the constraints on the EFT of inflation imposed by the symmetry were considered in \cite{DBIInflation1,DBIInflation2}.

 
\subsection{Degree two}\label{ssec:}

Once commuted with translations, transformations of degree two must give a transformation of degree one that is also a symmetry. If this is the DBI generator \eqref{DBIsymmetry}, the theory must take the form of the DBI action. But the DBI action has no free parameters to adjust and, by direct inspection, it is not invariant under any transformation of degree two. We conclude that the relevant transformations of degree one are Lorentz transformations and the scaling symmetry \eqref{Scaling-Symmetry}. By applying a similar analysis to above we see that a $P(X)$ theory which is invariant under an $N=2$ transformation must be a subset of \eqref{Scaling-Superfluid} and requires $\alpha = D/2$ i.e.~$\Delta=-a_{3} = 0$. The new symmetry is generated by a vector $K_{\mu}$ and is of the form
\begin{align}
\delta_{K_{\mu}} \phi = x^{2} \partial_{\mu}\phi - 2 x_{\mu}x^{\nu}\partial_{\nu}\phi. \label{SpecialConformalSymmetry}
\end{align}
This symmetry commutes with $Q$ and forms the $SO(2,D)$ conformal algebra with the other generators, i.e.~$K_{\mu}$ is the generator of special conformal transformations. The leading order invariant is the \textit{Conformal Superfluid}
\begin{align}
S = \int d^{D}x X^{D/2}\,, \label{Conformal-Superfluid}
\end{align} 
which has been studied in e.g.~ \cite{ConformalSuperfluid1, ConformalSuperfluid2, ConformalSuperfluid3, ConformalSuperfluid4}. We remind the reader that the non-zero commutators of the conformal algebra, in addition to \eqref{Poincare-Algebra}, are
\begin{align}
[M_{\mu\nu},K_{\sigma}] &= 2\eta_{\sigma [\mu}K_{\nu]}, & [K_{\mu},\hat{D}] &= -K_{\mu} \\ 
 [P_{\mu}, \hat{D}] &= P_{\mu}, & [P_{\mu},K_{\nu}] &= 2M_{\mu\nu} + 2 \eta_{\mu\nu} \hat{D}.\nonumber
\end{align}
 It is interesting to note that the extension of \eqref{Scaling-Superfluid} to full conformal symmetries is only possible when $\Delta = 0$. Indeed if we take the conformal group and augment it with the commutator $[Q,D] = -\Delta Q$ we see that the Jacobi identities are only satisfied when $\Delta = 0$. This can also be seen at the level of the transformation rules. For $\Delta \neq 0$ the transformation rule \eqref{SpecialConformalSymmetry} would need an additional term of the form $\phi x_{\mu}$ but then the commutator $[Q,K_{\mu}]$ would generate another vector transformation (equation \eqref{DBIsymmetry} with $a_{4} = 0$) which is only a symmetry when \eqref{Superfluid-action} reduces to a free theory.

It is simple to understand why the Conformal Superfluid is conformally invariant even though $\phi$ is not the dilaton (apart from the fact that the trace of its energy-momentum tensor vanishes!). Consider the $D$-dimensional theory of a dilaton $\psi$ which can be constructed out of diffeomorphism invariant combinations of the metric $g_{\mu\nu} = \psi^{2} \eta_{\mu\nu}$ and is a non-linear realisation of $SO(2,D)$. Now if we wish to couple a shift-symmetric scalar field $ \chi $ with scaling dimension $ \Delta_{\chi}=0$ to the dilaton in such a way that full conformal symmetries remain intact we again write down diffeomorphism invariant actions using the $g_{\mu\nu}$ metric \cite{Sundrum}. So if we wish to couple a matter field $\chi$ with a $P(X)$ type action would we have
\begin{align}
S_{\text{matter}} = \int d^{D}x \psi^{D} P(\psi^{-2} \eta^{\mu\nu} \partial_{\mu}\chi \partial_{\nu} \chi) \,,
\end{align}
and then it is simple to see that the choice $P = X^{D/2}$ is the only case where the dilaton drops out. So we are left with a conformally invariant theory in the absence of the dilaton and all symmetries of the conformal group act linearly on the matter field $\chi$, or equivalently, $\phi$. Indeed in the passive form of the transformation rules $\phi$ does not transform under dilatations (since in this case $\Delta = 0$) or special conformal transformations and therefore its active transformation rules presented above are solely due to its spacetime dependence. In any case, let us emphasise that these symmetries do indeed act non-linearly on the phonon $\pi$, as desired.

 
\subsection{Degree three and higher}\label{DegreeThreeHigher}

Since the Conformal Superfluid has no free coupling constants or parameters, it cannot be constrained further by additional symmetries. However, let us check whether the theory \eqref{Conformal-Superfluid} already possesses additional higher degree conserved currents. We will prove a general theorem that restricts the extensions of the Conformal group in $  D\geq 3$. In particular, if special conformal transformations are the only degree two symmetries the conformal group in $  D\geq 3 $ cannot be extended with any symmetry of degree $  N=3 $ or higher. The proof relies only on the non-existence of the algebra and is therefore valid for any Conformal Field Theory obeying the assumption above. For the specific case of the Conformal Superfluid, any new symmetries must be of degree three or higher since we have classified all lower degree possibilities. The above theorem then tells us that our classification of symmetric superfluids is already complete at degree two and we collect all generators discussed above in Table \ref{tab2}.

First consider a degree three transformation (where terms can have at most three powers of the coordinates) generated by some generator $S^{3}_{m}$. We know that the commutator between translations and $S^{3}_{m}$ must give back special conformal transformations $K_{\mu}$ since this is the only degree two generator. We therefore have 
\begin{align}
[P_{\mu},S^{3}_{m}] = t_{\alpha \mu\sigma}K^{\sigma} + S^{1}_{m+1}\label{DegreeThree}
\end{align}
where $  \alpha=\{\alpha_{1}\dots \alpha_{m}\} $, $S^{1}_{m+1} $ represents symmetries of degree one and zero (Lorentz, dilatations, translations, shift symmetry) which won't play a role in what follows, and $t_{\alpha \mu \sigma}$ is constructed out of Lorentz invariant quantities, namely $  \eta_{\mu\nu} $ and the Levi-Civita symbol $  \epsilon_{\mu_{1}\dots \mu_{D}} $.

Now consider the Jacobi identity involving two copies of $P_{\mu}$ and one copy of $S^{3}_{m}$. Since $[P_{\mu},P_{\nu}] = 0$ there are two contributions which combine into
\begin{align}\label{Ji}
[P_{[\mu},[P_{\nu]},S^{3}_{m}]] = t_{\alpha[\nu|\sigma|}[P_{\mu]}, K^{\sigma}] = 0.
\end{align}
From the $SO(2,D)$ conformal algebra we know that 
\begin{align}
[P_{\mu},K_{\nu}] = 2 M_{\mu\nu} + 2 \eta_{\mu\nu} \hat{D}
\end{align}
and therefore the Jacobi identity requires
\begin{align}
 t_{\alpha[\nu|\sigma|}(M_{\mu]}{}^{\sigma} + \delta_{\mu]}^{\sigma} \hat{D}) = 0.
\end{align}
For the coefficient of $\hat{D}$ to vanish we require 
\begin{align}
 t_{\alpha[\nu \mu]} = 0
\end{align}
i.e.~$t$ is symmetric in its last two indices, however this is not enough to kill the coefficient of $M_{\mu\sigma}$. To do so we need to place more constraints on $t$. 

We have
\begin{align}
2t_{\alpha [\nu \sigma} M_{\mu]}{}^{\sigma} =t_{\alpha \nu \sigma} M_{\mu}{}^{\sigma} -t_{\alpha \mu \sigma} M_{\nu}{}^{\sigma} =0\,,
\end{align}
which is non-trivial only for $  \mu \neq \nu $. Let us write out the sum over $  \sigma $ for some fixed values of $  \mu=\bar\mu\neq \nu=\bar \nu $:
\begin{align}
t_{\alpha \bar \nu \bar \nu} M_{\bar \mu}{}^{\bar \nu} +t_{\alpha \bar \nu \rho} M_{\bar \mu}{}^{\rho} =t_{\alpha \mu\bar  \mu} M_{\bar \nu}{}^{\mu} +t_{\alpha \bar \mu \sigma} M_{\bar \nu}{}^{\sigma}
\end{align}
for $\rho,\sigma \neq \bar \mu,\bar \nu$ and where we do not sum over barred indices. Since $  \bar \mu\neq \bar \nu $ and $   \rho,\sigma\neq\bar  \mu,\bar \nu $, the following terms need to cancel separately:
\begin{align}
t_{\alpha \bar \nu\bar  \nu} M_{\bar \mu}{}^{\bar \nu} &=t_{\alpha \bar \mu \bar \mu} M_{\bar \nu}{}^{\bar \mu} \,, \label{weak}\\
t_{\alpha \bar \nu \rho} M_{\bar \mu}{}^{\rho}&=0 \hspace{0.1cm}  \Rightarrow \hspace{0.1cm} t_{\alpha \bar \nu \rho}=0 \quad (\rho\neq \bar \nu, \bar \mu;\, D\geq 3) \,.  \label{strong}
\end{align}
It is clear that the second of these constraints does not apply in $  D=2 $ since then it is not possible to have three distinct indices $  \{\rho,\bar \nu,\bar \mu\} $. These constraints therefore need to be studied separately in $  D=2 $ and $  D\geq 3 $.

We first start with $D \geq 3$ where from \eqref{strong} we know that $  t_{\alpha\bar  \nu \rho} $ vanishes for every $ \bar  \nu \neq \rho $, so we are only left with discussing $  t_{\alpha \bar \nu\bar \nu} $. We consider separately the cases with $  \bar \nu=0 $ and $  \bar \nu $ spatial. For $ \bar  \nu=0\neq \bar \mu $ we notice that $  M_{\bar \mu}{}^{\bar \nu}=+M_{\bar \nu}{}^{\bar \mu} $ and so from \eqref{weak} we conclude that
\begin{align}
t_{\alpha 00}=t_{\alpha ii}\,,\label{near}
\end{align}
for $  i $ a spatial index. For both $ \bar  \nu $ and $ \bar  \mu $ spatial (but different) instead $  M_{\bar \mu}{}^{\bar \nu}=-M_{\bar \nu}{}^{\bar \mu} $ and so 
\begin{align}\label{near2}
t_{\alpha ii}=-t_{\alpha jj}
\end{align}
for any $  i\neq j $ with both $  i $ and $  j $ spatial. This constraint admits a non-trivial solution only for $  D=3 $, while for $  D\geq 4 $ the only solution is $  t_{\alpha ii}=0 $ for any spatial $  i $. Either way, for any $  D\geq 3 $, the combination of \eqref{near} and \eqref{near2} imposes $  t_{\alpha 00 }=-t_{\alpha 00} $ and so $  t_{\alpha \bar \nu\bar \nu} $ also vanishes for any $  \bar \nu $. Since these constraints hold for any $\bar{\mu}$ and $\bar{\nu}$, we have proven that $  t_{\alpha\mu\nu} $ vanishes for any value of its indices $  \alpha, \mu $ and $  \nu $ and so no symmetry of degree 3 or higher is allowed in $  D\geq 3 $: its existence is incompatible with Jacobi identities. Notice that in the proof we did not have to assume anything about how $  t_{\alpha\mu\nu} $ is constructed.

Now as expected\footnote{It is well known that in $D=2$ the conformal group can be extended by infinitely many additional generators \cite{CFT}.} $ D=2 $ is special since \eqref{strong} does not apply. Here the only constraint we get is for $  \bar \mu =1 $ and $  \bar \nu=0 $, for which $  M_{0}{}^{1}=M_{1}{}^{0} $ and so from \eqref{weak}
\begin{align}
t_{\alpha 00}=t_{\alpha 11}\quad \Leftrightarrow \quad t_{\alpha \mu\nu}\eta^{\mu\nu}=0\,.
\end{align}
This condition is not strong enough to kill $t$ since tensors with the above property certainly exist. To push further, one should further demand that $  t_{\alpha\mu\nu} $ is built out of Lorentz invariant tensors, namely the Minkowski metric $  \eta_{\mu\n} $ and the Levi-Civita symbol $  \epsilon_{\mu\nu} $. But even this constraint doesn't rule out all degree 3 symmetries and in fact, using the notation of \cite{Blumenhagen:2009zz}, the generators $  l_{2}=-z^{3}\partial_{z} $ and $  \bar l_{2}=-\bar z^{3} \partial_{\bar z}$ provide precisely an example of an $  S_{4}^{3} $ symmetry in $  D=2 $. However, the only theory which could in principle be invariant under higher degree symmetries is the Conformal Superfluid, which in $D=2$ becomes a free theory. As we have already mentioned, the free theory is actually invariant under an infinite number of symmetries in any number of dimensions. 

In conclusion, any Conformal Superfluid with self-interactions is not invariant under any symmetries beyond $SO(D,2) \times U(1)$. 


\subsection{Tadpoles and driven superfluids} \label{Tadpoles}

In the absence of gravity, the shift symmetry allows also for a non-trivial potential 
\begin{align}
V(\phi)=\lambda \phi
\end{align} 
for some coupling constant $  \lambda $, which shifts by a total derivative (see e.g.~ \cite{Finelli:2018upr}). 
In the absence of this linear potential, all superfluids admit the Lorentz breaking vacuum solution $\langle \phi(x) \rangle = \mu t $, around which time translations and shifts are non-linearly realised while a linear combination of the two is linearly realized. This is the prototypical mechanism of spontaneous symmetry probing discussed in \cite{Nicolis:2011pv}, where one can still define a linearly realized modified Hamiltonian and perturbations can be parameterized in such a way that their Lagrangian is manifestly time-translation invariant. On the other hand, in the presence of the linear potential, a general homogenous solution for $  \phi(t) $ is non-linear in $  t $ and perturbations live in a truly time dependent background. As long as we expand around an exact background solution $  \phi(t) $, the action for perturbations start quadratic as usual and the theory can be quantized without any problematic tadpole terms.

Let us check which symmetries are consistent with this term. We need to require that $\delta \phi$ itself is a total derivative. This is the case for all the vector symmetries in \eqref{SquareRootSymmetry}, the DBI symmetry \eqref{DBIsymmetry} and the scaling symmetry \eqref{Scaling-Symmetry} when $\Delta = D$. Since closure of the algebra requires $\Delta = 0$ for the Conformal Superfluid, the tadpole is not permitted in this case.

 
\subsection{$  D=1 $: Quantum mechanics }\label{ssec:QM}

Before ending this section and studying the Scaling Superfluid in more detail, let us first briefly discuss the $D=1$, quantum mechanical, case which is special\footnote{We are thankful to Bernardo Finelli for pointing out to us the special nature of $  D=1 $ and for describing to us the example discussed here.}. In this case, there are no Lorentz symmetries and therefore no Lorentzian distinction between different generators with $ m\neq m'$. This allows one, for example, to take linear sums of generators which in $D \neq 1$ would live in different Lorentz representations. For example, consider the DBI symmetry \eqref{DBIsymmetry} and the scaling symmetry \eqref{Scaling-Symmetry}. For all $D$, there are no theories which are invariant under both of these transformations simultaneously. However, in $D=1$ we can find a linear combination of these two transformations that is a symmetry of a theory that is not invariant under the individual transformations. In $D=1$ these symmetries respectively act on $\phi$ as
\begin{align}
\delta_{A_{0}}\phi = -t + \phi \dot{\phi}, \quad \delta_{\hat{D}}\phi = -\Delta \phi - t \dot{\phi}
\end{align}
and on $\pi$ as
\begin{align}
\delta_{A_{0}}\pi &= (\mu^{2}-1)t + \mu \pi + \mu t \dot{\pi} + \pi \dot{\pi} \\ \delta_{\hat{D}}\pi &= -\mu(1 + \Delta) t - \Delta \pi - t \dot{\pi}.
\end{align}
It is possible to take a linear combination of these two transfomation rules such that the $t \dot{\pi}$ term drops out. One can then find a non-trivial invariant of the resulting symmetry \cite{BFSymmetry}. Any exhaustive classification of symmetries in $D=1$ therefore requires further analysis.

\section{Scaling Superfluids}\label{sec3}

Let's take a closer look at the Scaling Superfluid action
\begin{align}
S = \int d^{D}x X^{\alpha} \,,\label{SuperfluidFluctuations}
\end{align}
where the symmetries act as follows
\begin{align}
\delta_{P_{\mu}}\phi &= -\partial_{\mu}\phi & &\text{(translations)}\,, \\
\delta_{J_{\mu\nu}}\phi &= 2x_{[\mu}\partial_{\nu]} \phi & &\text{(Lorentz)}\,, \\
 \delta_{Q} \phi &=1 & &\text{(shift)}\,,\\
 \delta_{\hat{D}}\phi &= \at \phi - x^{\mu}\partial_{\mu}\phi & &\text{(scaling)}\,,
\end{align}
where  
\begin{align}
\Delta=\frac{D}{2\alpha}-1\,. 
\end{align}
The associated Noether currents are
\begin{align}
J_{Q}^{\mu}&\equiv -2\alpha X^{\alpha-1} \partial^{\mu}\phi\,, \\
J_{\hat{D}}^{\mu}&\equiv -2\alpha X^{\alpha-1} \partial^{\mu}\phi \left(  \at\phi -x^{\nu}\partial_{\nu}\phi\right)+x^{\mu} X^{\alpha}\label{scaling}\\
T_{\mu\nu}&=2\alpha X^{\alpha-1}\partial_{\mu}\phi\partial_{\nu}\phi+g_{\mu\nu}X^{\alpha} \,.
\end{align}
The classical equations of motion 
\begin{align}
\partial_{\mu}\left(  X^{\alpha-1}\partial^{\mu} \phi\right)=0\label{eom}
\end{align}
are exactly equivalent to the conservation of the shift-symmetry current $ \partial_{\mu}J^{\mu}_{Q}=0$. Finally, for future reference we quote some mass-dimensions 
\begin{align}
[\phi] &= M^{D/(2\alpha)-1} & [X]&=M^{D/\alpha}  \\
[J_{Q}^{\mu}]&=M^{D-D/(2\alpha)} &  [J_{\hat{D}}^{\mu}]&=M^{D-1} \\
[T_{\mu\nu}] &= M^{D}
\end{align}

 
\subsection{Scaling without conformal symmetry}

It is a general result (see e.g \cite{Coleman:1970je,Callan:1970ze,Polchinski}) that the scaling current takes the form
\begin{align}
J_{\hat{D}}^{\mu}=x^{\nu}T_{\nu}^{\mu}+K^{\mu}\,,
\end{align}
where the virial current $ K^{\mu} $ is constructed with just local operators (i.e.~no explicit $x^{\mu}$ dependence). Conservation of this current on the solution of the equations of motion implies
\begin{align}
T^{\mu}_{\mu}=-\partial_{\mu}K^{\mu}.
\end{align}
On the other hand, a conformal transformation is defined (in the passive form) by the change of coordinates $ \delta x^{\mu}=v^{\mu}(x) $ with
\begin{align}
2\partial_{(\mu}v_{\nu)}=\frac{2}{D} \eta_{\mu\nu}\partial_{\rho}v^{\rho}\,,
\end{align}
where $ \partial_{\mu}v^{\nu} $ spans all affine functions of $ x $ (linear plus constant). Conformal symmetry leads to the current
\begin{align}
J^{\mu}_{v}=v^{\nu}T_{\nu}^{\mu}+\partial_{\rho}v^{\rho}K'^{\mu}+\partial_{\nu}\partial_{\rho}v^{\rho}L^{\nu\mu}\,,
\end{align}
with $ L^{\nu\mu} $ also made of local operators and $ K'^{\mu} $ being the same as $ K^{\mu} $ up to a conserved current. Conservation of the conformal current on the solution of the equations of motion implies (by matching order by order in $ x $)
\begin{align}
T^{\mu}_{\mu}=\partial_{\mu\nu}L^{\mu\nu}\,.
\end{align}
In this case, one can improve $  T_{\mu\nu} $ to some $  \theta_{\mu\nu} $ that is traceless $  \theta^{\mu}_{\mu}=0 $. So a necessary condition to have scaling but not conformal symmetry is the existence of a current $ K^{\mu} $ that is not conserved $  \partial_{\mu}K^{\mu}\neq 0 $ and is not a total derivative. In scalar theories which are perturbative around the Poincar\'{e} invariant vacuum this cannot happen because the only dimension $ D-1 $ current one can write is $ \partial_{\mu}(\phi)^{2} $, which is a total derivative  \cite{Polchinski}. Let us see at what happens in our Scaling Superfluid. The scaling current in \eqref{scaling} can be re-written as 
\begin{align}
J_{\hat{D}}^{\mu}&=x^{\nu}T_{\nu}^{\mu}+\Delta \alpha X^{\alpha-1} \partial^{\mu}\left( \phi \right)^{2}\\
&=x^{\nu}T_{\nu}^{\mu}-2 \Delta \phi J^{\mu}_{Q}\,.
\end{align}
Since the theory is not perturbative around $\langle \phi(x) \rangle =\text{const} $, the field $ \phi $ does not have dimension $ (D-2)/2 $ and one can build a current $ K^{\mu} $ of dimension $ D-1 $ that is not a total derivative by multiplying $ \partial_{\mu}\phi^{2} $ with the appropriate power of $ X $:
\begin{align}
K^{\mu}=-2 \Delta \phi J^{\mu}_{Q}=\Delta \alpha X^{\alpha-1} \partial^{\mu}\phi^{2}\,.
\end{align}
This virial current indeed vanishes for Conformal superfluids, since then $  \Delta=0 $. Because the virial is not a total derivative, the stress tensor of Scaling Superfluids cannot be improved to become traceless. Indeed
\begin{align}
T^{\mu}_{\mu}=X^{\alpha}\left[  D-\frac{D}{(1+\Delta)}\right]\,,
\end{align}
and, as expected, it vanishes only for $ \Delta=0 $, $ \alpha=D/2 $.


 
\subsection{Perturbation theory}\label{ssec:}

Let's redefine $\phi$ to be dimensionless and study the Scaling Superfluid action 
\begin{align}
S = \int d^{D}x \mu^{D} \left(\frac{X}{\mu^{2}} \right)^{\alpha} \,,\label{SuperfluidFluctuations}
\end{align}
around the homogeneous and isotropic background $\phi = \mu t + \pi$, as relevant for cosmology. The scale symmetry acts non-linearly on $\pi$ as
\begin{align}
\delta_{\hat{D}} \pi = -\mu t(\Delta+1) -\Delta \pi- x^{\mu}\partial_{\mu}\pi.
\end{align}
When $\Delta = 0$ there are further symmetries, special conformal transformations, which act on $\pi$ as
\begin{align}
\delta_{K_{0}} \pi &= (t^{2} + x^{i}x_{i})(\mu + \dot{\pi}) + 2 t x^{i}\partial_{i}\pi\,, \\
\delta_{K_{i}} \pi &= (x^{j}x_{j} - t^{2}) \partial_{i}\pi - 2t x_{i}(\mu + \dot{\pi}) - 2 x_{i}x^{i}\partial_{j}\pi.
\end{align}
To exhibit the details of the $\pi$ EFT we expand \eqref{SuperfluidFluctuations} up to quartic order which, after canonically normalising, yields
\begin{align}\label{lag}
S = n \int d^{D}x \left(\frac{1}{2} \dot{\pi}^{2} - \frac{c_{s}^{2}}{2} \partial^{i}\pi \partial_{i}\pi + \frac{g_{1}}{\mu^{D/2}}\dot{\pi}^{3} \nonumber \right. \\ \left. + \frac{g_{2}}{\mu^{D/2}} \dot{\pi} \partial^{i}\pi \partial_{i}\pi + \frac{g_{3}}{\mu^{D}} \dot{\pi}^{4} \nonumber \right. \\ \left. + \frac{g_{4}}{\mu^{D}} \dot{\pi}^{2} \partial^{i}\pi \partial_{i}\pi + \frac{g_{5}}{\mu^{D}} (\partial^{i}\pi \partial_{i}\pi)^{2} \right)\,,
\end{align}
where 
\begin{align}
c_{s}^{2} = \frac{1+\Delta}{d -\Delta}, \qquad n = \frac{1}{c_{s}^{2}}\left(\frac{1}{c_{s}^{2}} + 1 \right)\,,
\end{align}
and 
\begin{align}
g_{1} &= \frac{1}{6 c_{s}^{2}}(1 - c_{s}^{2}) \,, & g_{2} &= -\frac{1}{2}(1 - c_{s}^{2}) \,,\\
g_{3} &= \frac{1}{24 c_{s}^{4}}(1 - c_{s}^{2})(1 - 2 c_{s}^{2}) \,, \\
g_{4} &= -\frac{1}{4 c_{s}^{2}}(1 - c_{s}^{2})(1 - 2 c_{s}^{2})\,, & g_{5} &= \frac{1}{8}(1 - c_{s}^{2})\,.
\end{align}
We note that when the couplings $g_{i}$ are arbitrary the action \eqref{lag} is the general decoupling limit of the effective field theory of shift symmetric, single clock cosmology to leading order in derivatives. This flat space limit is valid for energies above the Hubble scale. For the Scaling Superfluid we see that all couplings for $\pi$ are fixed in terms of the speed of sound $c_{s}^{2}$ which in turn is fixed in terms of $\Delta$ once the spacetime dimension is specified. This is therefore a very predictive theory of single clock cosmology, on the same footing as the DBI case \cite{DBIInflation2}. We plan to investigate the cosmology of the Scaling Superfluid in more detail in future work.

Let us collect the constraints on $ \alpha$ to avoid gradient instabilities, superluminality and ghosts\footnote{See also \cite{Imaginary} for a discussion on inflationary EFTs with an imaginary speed of sound ($c_{s}^{2} < 0$).}. We respectively have 
\begin{align} 
c_{s}^{2}\geq 0  &\then   \alpha > \frac{1}{2}\\ 
c_{s}^{2}\leq 1  &\then   \alpha < \frac{1}{2} \quad \text{or} \quad  \alpha \geq1\\ 
n>0 &\then \alpha<0 \quad \text{or} \quad \alpha >\frac{1}{2}\,.
\end{align}
These can be collectively satisfied for
\begin{align}
\alpha \geq1\Leftrightarrow -1<\Delta<\frac{D}{2}-1 \quad \text{(no pathologies)}  \,.
\end{align}
It is interesting to compute the cutoff of this theory. We can use perturbative unitarity of the 2-to-2 scattering of $ \pi $, and find that the strong coupling cutoff $\Lambda$, in $D=4$, is given by \cite{Signs}
\begin{align}
\left(\frac{\Lambda}{\mu}\right)^{4}=\frac{30 \pi (c_{s}^{3} + c_{s}^{5})}{\left( 1-c_{s}^{2} \right)^{2}}\,.
\end{align}
Above $\Lambda$ loop corrections dominate scattering amplitudes and the theory is strongly coupled. As $ \mu \rightarrow 0$, the strong coupling scale also goes to zero, which matches our expectation that the Poincar\'e invariant background does not allow for a perturbative description. One might want to take the double limit $ \mu \rightarrow 0 $ and $ c_{s}^{2}\rightarrow 1 $ in such a way to keep $ \Lambda $ finite. Indeed this is possible, but it simply gives the free theory, since $\alpha(c_{s}=1)=1$\footnote{It is also interesting to note that, by computing the $2$-to-$2$ scattering amplitude using the results of \cite{Signs}, we see that the conditions for stability of the theory (which boil down to $\alpha \geq 1$) are stronger than those imposed by demanding positivity of the amplitude in the forward limit. This was also seen in \cite{EnergyAmplitude} in the context of the Conformal Galileon which is a non-linear realisation of the four-dimensional conformal group. A possible resolution may come from postivity constraints associated with higher point scattering amplitudes and from non-forward amplitudes.}. 

Finally, let us discuss the Renormalization Group flow of this theory. This (effective) theory hits strong coupling at the scale $ \Lambda $ and so we cannot say much about its UV fixed point. But the IR fixed point around $\langle \phi(x) \rangle =\mu t $ is under perturbative control. Indeed, as we decrease energy all interactions in \eqref{lag} become weaker and weaker such that the theory eventually becomes free and therefore fully conformal in the deep IR. We note that this behaviour is not guaranteed when Lorentz symmetry is spontaneously broken, see e.g.~ \cite{GhostCondensate, Fermi}.
%

 
\subsection{Higher order operators}\label{ssec:}

We now consider the leading corrections to \eqref{lag} which have more derivatives on $\pi$ under the assumption that the symmetries of the classical Lagrangian remain valid at the quantum level. We plan to investigate possible quantum anomalies in future work. For the Scaling Superfluid it is simple to compute these corrections: all interactions are fixed by counting fields and derivatives. However for the Conformal Superfluid it is more complicated since the interactions must also be invariant under special conformal transformations. So in that case we turn to the coset construction.

We compute the corrections in terms of $\phi$. Initially consider a correction linear in $\partial \partial \phi$ i.e.~a correction of the form
\begin{align}
F(X) \Box \phi\,. \label{LinearInBox}
\end{align}
Note that the only other possible index contraction can be brought to this form by integration by parts. Now by demanding invariance under \eqref{Scaling-Symmetry} we require
\begin{align}
F(X) = X^{\frac{D-2 -\Delta}{2(1+\Delta)}}\,.
\end{align}
More generally if we have a correction which is non-linear in $\partial \partial \phi$ the symmetry fixes it to take the schematic form
\begin{align}
X^{\frac{D - 2m -\Delta m}{2(1+\Delta)}} (\partial \partial \phi)^{m}\,. \label{GeneralCorrection}
\end{align}
The coefficents of these higher derivative corrections are not fixed by the scaling symmetry, they are therefore free coupling constants which must be fixed by matching to a (partial) UV completion or by observations. Furthermore, to stay within the regime of validity of the Scaling Superfluid EFT, these higher order corrections must always be considered perturbatively, see e.g.~ \cite{Vainshtein}.

Now for the Conformal Superfluid we use the coset construction to find the leading order corrections. The relevant symmetry breaking pattern corresponds to the group $SO(D,2) \times U(1)$, with generators $P_{\mu}, M_{\mu\nu}, \hat{D}, K_{\mu}$ and $Q$, broken down to a subgroup consisting of spatial rotations $M_{ij}$, spatial translations $\bar{P}_{i}$ and a new form of time translations given by $\bar{P}_{0} = P_{0} + \mu Q$. At the level of the algebra this change of basis ensures that there are inverse Higgs constraints \cite{InverseHiggs} which enable us to eliminate the $d$ Goldstones $\eta^{i}$ associated with the broken boosts since $[\bar{P}_{i},M_{0j}] \supset \delta_{ij} \mu Q$, and the dilaton $\psi$ since $[\bar{P}_{0}, \hat{D}] \supset \mu Q$. As is well known, the Lorentzian vector $\tau^{\mu}$ associated with special conformal transformations $K_{\mu}$ can also be eliminated by an inverse Higgs constraint since $[\bar{P}_{\mu},K_{\mu}] \supset \eta_{\mu\nu}\hat{D}$. The full symmetries can therefore by realised on $\pi$ which is the Goldstone of the spontaneously broken $U(1)$. For notational convenience we have taken $\bar{P}_{i} = P_{i}$. For a discussion on the conditions which must be met by the algebra such that the necessary inverse Higgs constraints exist and more generally for an introduction to the coset construction we refer the read to \cite{HigherOrderIH}. 

We parametrise the coset element as 
\begin{align}
\Omega = e^{x^{\mu}\bar{P}_{\mu}} e^{\pi Q}e^{\eta^{i}M_{0i}}e^{\psi \hat{D}}e^{\tau^{\mu}K_{\mu}}\,,
\end{align}
which yields the Maurer-Cartan (MC) form
\begin{align}
&\Omega^{-1} \partial_{\mu} \Omega = e^{\psi} \Lambda^{\nu}{}_{\mu} \bar{P}_{\nu}  + (\omega^{ij}_{M_{ij}})_{\mu}M_{ij}  + (\omega_{\hat{D}})_{\mu}\hat{D}  \\ \nonumber +  &  (\omega^{i}_{M_{0i}})_{\mu}M_{0i} +  (\omega_{K_{0}})_{\mu}K_{0} + (\omega^{i}_{K_{i}})_{\mu}K_{i} + (\omega_{Q})_{\mu}Q\,,
\end{align}
where the MC components are given by
\begin{align}
(\omega^{ij}_{M_{ij}})_{\mu} &= (1-\gamma) \partial_{\mu}\beta^{[i} \beta^{j]} / \beta^{2} + 2 e^{\psi} \Lambda^{[i}{}_{\mu}\tau^{j]}\,, \\
(\omega_{\hat{D}})_{\mu} &= (\partial_{\mu} \psi  + 2 e^{\psi}\Lambda^{\nu}{}_{\mu}\tau_{\nu}) \,, \\
(\omega^{i}_{M_{0i}})_{\mu} &=  - (\gamma \Lambda^{i}{}_{j}\partial_{\mu}\beta^{j} + 4e^{\psi}\Lambda^{[i}{}_{\mu}\tau^{0]}) \,,\\
(\omega_{K_{0}})_{\mu} &= \partial_{\mu} \tau^{0} + \tau^{0} \partial_{\mu} \psi  + 2 e^{\psi} \tau^{0}\Lambda^{\nu}{}_{\mu}\tau_{\nu} \nonumber \\ & - e^{\psi} \tau^{2}\Lambda^{0}{}_{\mu} + \gamma \tau_{i}\Lambda^{i}{}_{j}\partial_{\mu}\beta^{j} \,,\\ 
(\omega^{i}_{K_{i}})_{\mu} & = \partial_{\mu} \tau^{i} + \tau^{i} \partial_{\mu} \psi  + 2 e^{\psi} \tau^{i}\Lambda^{\nu}{}_{\mu}\tau_{\nu} - e^{\psi} \tau^{2}\Lambda^{i}{}_{\mu}  \nonumber \\  & + \gamma \tau^{0}\Lambda^{i}{}_{j}\partial_{\mu}\beta^{j} - 2(1 - \gamma)\partial_{\mu}\beta^{[i}\beta^{j]}\tau_{j} / \beta^{2})  \,,\\
(\omega_{Q})_{\mu} & =  \partial_{\mu} \pi + \mu (\delta_{\mu}^{0} - e^{\psi} \Lambda^{0}{}_{\mu})\,.
\end{align} 
In arriving at the above, we have made the field redefinition
\begin{align}
\beta^{i} = - \eta^{i}\frac{\tanh \sqrt{\eta^{2}}}{\sqrt{\eta^{2}}} \,,\label{BoostRedef}
\end{align}
and defined
\begin{equation}
\Lambda = \left(\begin{array}{cc} \gamma & - \gamma \vec{\beta}\\ -\gamma \vec{\beta} & ~ \delta^{ij} + \frac{\gamma - 1}{\beta^{2}}\beta^{i} \beta^{j} \end{array}\right) \,, \quad \gamma = \frac{1}{\sqrt{1 - \beta^{2}}}\,.
\end{equation}
Note that this is the \textit{exact} MC form for this symmetry breaking pattern. 

We remind the reader that the MC components along $\bar{P}_{\mu}$ are interpreted as vielbeins $e^{\mu}{}_{\nu}$, so in this case we have $e^{\mu}{}_{\nu} = e^{\psi} \Lambda^{\mu}{}_{\nu}$, which enables us to define a metric and invariant measure
\begin{align} 
g_{\mu\nu} = e^\rho{}_\mu e^\sigma{}_\nu \eta_{\rho\sigma}, \qquad \sqrt{-g}d^D x.
\end{align} 
Furthermore, for each Goldstone we have a corresponding covariant derivative. For example, the $\pi$ covariant derivative is given by
\begin{align}
\nabla_{\mu} \pi = (e^{-1})^{\nu}{}_{\mu}(\omega_{Q})_{\nu}. \label{CovariantDeriv}
\end{align}
Invariant theories are then constructed out of the covariant derivatives, with the indices contracted in an $SO(d)$ invariant manner, integrated over spacetime with the above invariant measure. The MC components along any unbroken generators, so in this case $(\omega^{ij}_{M_{ij}})_{\mu}$, are used to couple matter fields to the Goldstones. 

Now as we discussed above, the fields $\beta^{i}$, $\psi$ and $\tau^{\mu}$ can be eliminated from the Goldstone EFT by inverse Higgs constraints. The relevant inverse Higgs constraints come from setting $\nabla_{i} \pi = 0$, $\nabla_{0}\pi = 0 $ and $\nabla_{\mu}\psi = 0$, respectively. By solving these constraints we find
\begin{align}
\beta_{i} &= - \frac{\partial_{i}\pi}{\mu + \dot{\pi}}, \\ e^{\psi} &= \frac{\sqrt{X}}{\mu}\,, \\
\tau_{\mu} &= -\frac{1}{2}e^{-\psi} \Lambda_{\mu}{}^{\nu}\partial_{\nu}\psi\,.
\end{align}
Since $\text{det}(\Lambda) = 1$, the leading order invariant is therefore
\begin{align}
S = q_{0}\int d^{D}x \mu^{D} e^{\psi D} = q_{0} \int d^{D}x X^{D/2} \,,
\end{align}
as expected. Here and in the following $q_{i}$ are arbitrary, dimensionless coupling constants which are not fixed by the conformal symmetry. Invariants with more derivatives are constructed using the covariant derivatives $\nabla_{\mu}\beta^{i}$ and $\nabla_{\mu}\tau_{\nu}$. First consider the leading order correction which is of the form \eqref{LinearInBox} which, by counting derivatives, must come from the invariant
\begin{align}
q_{1}\int d^{D}x \mu^{D-1} e^{\psi D} \nabla_{i} \beta^{i}. \label{FirstCorrection}
\end{align}
By computing \eqref{FirstCorrection} on the inverse Higgs solutions we see that it is actually a total derivative and therefore there is no invariant of the form \eqref{LinearInBox}. If an operator of this form is to exist it must be a Wess-Zumino term (see e.g.~ \cite{SuperfluidWess} for a discussion on Wess-Zumino terms for Superfluids). However, we have checked explicitly using the transformation rules \eqref{Scaling-Symmetry} and \eqref{SpecialConformalSymmetry} that there is no correction to the Conformal Superfluid of this type. 

We now turn to corrections of the form \eqref{GeneralCorrection} with $m=2$, and of course $\Delta= 0$, which must come from terms quadratic in $\nabla \beta$ or linear in $\nabla \tau$. A priori there are a number of different contributions and here we do not study these corrections in full generality since that would also require an analysis of Wess-Zumino terms which is beyond the scope of this work but an interesting avenue for future work. However, we found that $\nabla_{0}\beta^{i} = 0$ on the inverse Higgs solution and an example of a simple invariant of this type is
\begin{align}
q_{2}\int d^{D}x \mu^{D-2} e^{\psi D} (\nabla_{i} \beta^{i})^{2}
\end{align}
where
\begin{align}
\mu^{D-2}e^{\psi D} (\nabla_{i} \beta^{i})^{2} =  \frac{(D-2)^{2}}{4}X^{\frac{D-8}{2}}(\partial_{\mu}\phi \partial^{\mu}X)^{2} \nonumber \\
+  X^{\frac{D-4}{2}} (\Box \phi)^{2} + (D-2)X^{\frac{D-6}{2}}\Box \phi \partial_{\mu}\phi \partial^{\mu}X.
\end{align}
Perhaps a more simple way to extract these $m=2$ invariants is to write down all independent derivative contractions which are consistent with the scaling symmetry \eqref{Scaling-Symmetry} then fix the relative coefficients by also demanding invariance under \eqref{SpecialConformalSymmetry}.

%
%


\section{Conclusions and outlook}

In this paper we have provided a full classification of all Superfluid actions \eqref{Superfluid-action} that display symmetries in addition to Poincar\'e and the shift symmetry, to lowest order in derivatives. Our results are summarised in Table \ref{tab}. There are only two possibilities: the well-known DBI action and Scaling Superfluids $  X^{\alpha} $. The Scaling Superfluid has additional conserved currents for specific choices of $\alpha$. For $\alpha = 1/2$ we found the additional symmetries \eqref{SquareRootSymmetry} and \eqref{additional} but we haven't shown whether or not even more exist. The presence of many conserved currents makes this theory somewhat peculiar and in our opinion merits further study. For $\alpha = D/2$ the scale symmetry is accompanied by invariance under special conformal transformations thereby leading to the Conformal Superfluid \eqref{Conformal-Superfluid}. For $D \neq 2$ i.e.~when this theory is interacting, we have shown that there are no additional symmetries: would-be higher degree symmetries are never compatible with Jacobi identities. We note that our results apply to any $P(X)$ theory since in our analysis we do not assume anything about the vacuum of the theory. 

There are many avenues for future research. While Poincar\'e symmetries are non-linearly realized on the Superfluid phonons $  \pi $, their action can be linearized by a field redefinition $  \phi=\mu t+\pi $. It would be interesting to find non-linear realizations of Poincar\'e on a single scalar that do not admit this linearization (as happens with additional degrees of freedom \cite{Bordin:2018pca}), or prove that they are impossible. It would also be nice to investigate the consequences of our results for cosmology, where for example Scaling Superfluids have already made an appearance (see e.g.~ \cite{Bruneton:2007si,Berezhiani:2015bqa,Khoury:2018vdv,Ferreira:2018wup} and references therein). Also, the Scaling Superfluid could be studied holographically, in analogy to Conformal Superfluids \cite{Herzog:2008he,Esposito:2016ria}. Finally, one should investigate what symmetries survive at the quantum level. Indeed one might worry that both the scaling and the conformal symmetries might be anomalous due to the necessary renormalization procedure, in the same way as for $  \lambda  \phi^{4} $. This should be easy to establish by checking the validity of the Ward-Takahashi identities for the scaling symmetry at one loop.



\section*{Acknowledgements}
We would like to thank Brando Bellazzini, Bernardo Finelli, Sebastian Garcia-Saenz, Garrett Goon, Tanguy Grall, Sadra Jazayeri, Remko Klein, Antonio Padilla, Riccardo Penco, McCullen Sandora, Diederik Roest, David Tong and Pelle Werkman for useful discussions. DS acknowledges the Dutch funding agency ``Netherlands Organisation for Scientific Research'' (NWO) for financial support. E.P. is partly supported by the research programme VIDI with Project No. 680-47-535, which is (partly) financed by NWO.



\begin{thebibliography}{99}

\bibitem{Coleman:1967ad}
  S.~R.~Coleman and J.~Mandula,
  ``All Possible Symmetries of the S Matrix,''
  Phys.\ Rev.\  {\bf 159} (1967) 1251.
  doi:10.1103/PhysRev.159.1251


\bibitem{Haag:1974qh}
  R.~Haag, J.~T.~Lopuszanski and M.~Sohnius,
  ``All Possible Generators of Supersymmetries of the S Matrix,''
  Nucl.\ Phys.\ B {\bf 88} (1975) 257.
  doi:10.1016/0550-3213(75)90279-5

\bibitem{Runaway}
  C.~P.~Burgess and M.~Williams,
  JHEP {\bf 1408} (2014) 074
  doi:10.1007/JHEP08(2014)074
  [arXiv:1404.2236 [gr-qc]].

\bibitem{Herzog:2008he} 
  C.~P.~Herzog, P.~K.~Kovtun and D.~T.~Son,
  Phys.\ Rev.\ D {\bf 79}, 066002 (2009)
  doi:10.1103/PhysRevD.79.066002
  [arXiv:0809.4870 [hep-th]].
  
\bibitem{Esposito:2016ria} 
  A.~Esposito, S.~Garcia-Saenz and R.~Penco,
  JHEP {\bf 1612}, 136 (2016)
  doi:10.1007/JHEP12(2016)136
  [arXiv:1606.03104 [hep-th]].
  
\bibitem{Afshordi:2006ad} 
  N.~Afshordi, D.~J.~H.~Chung and G.~Geshnizjani,
  Phys.\ Rev.\ D {\bf 75}, 083513 (2007)
  doi:10.1103/PhysRevD.75.083513
  [hep-th/0609150].
  
  \bibitem{Afshordi:2007yx} 
  N.~Afshordi, D.~J.~H.~Chung, M.~Doran and G.~Geshnizjani,
  Phys.\ Rev.\ D {\bf 75}, 123509 (2007)
  doi:10.1103/PhysRevD.75.123509
  [astro-ph/0702002].
  
 


\bibitem{ScaleVConformal1}
  Y.~Nakayama,
  ``Scale invariance vs conformal invariance,''
  Phys.\ Rept.\  {\bf 569} (2015) 1
  doi:10.1016/j.physrep.2014.12.003
  [arXiv:1302.0884 [hep-th]].
  
\bibitem{Polchinski}
  J.~Polchinski,
  ``Scale and Conformal Invariance in Quantum Field Theory,''
  Nucl.\ Phys.\ B {\bf 303} (1988) 226.
  doi:10.1016/0550-3213(88)90179-4

\bibitem{Dorigoni:2009ra}
  D.~Dorigoni and V.~S.~Rychkov,
  ``Scale Invariance + Unitarity => Conformal Invariance?,''
  arXiv:0910.1087 [hep-th].
  
\bibitem{Jackiw:2011vz}
  R.~Jackiw and S.-Y.~Pi,
  ``Tutorial on Scale and Conformal Symmetries in Diverse Dimensions,''
  J.\ Phys.\ A {\bf 44} (2011) 223001
  doi:10.1088/1751-8113/44/22/223001
  [arXiv:1101.4886 [math-ph]].
  
  
\bibitem{ScaleVConformal3}
  M.~A.~Luty, J.~Polchinski and R.~Rattazzi,
  ``The $a$-theorem and the Asymptotics of 4D Quantum Field Theory,''
  JHEP {\bf 1301} (2013) 152
  doi:10.1007/JHEP01(2013)152
  [arXiv:1204.5221 [hep-th]].
  
  
\bibitem{Fortin:2012hn}
  J.~F.~Fortin, B.~Grinstein and A.~Stergiou,
  ``Limit Cycles and Conformal Invariance,''
  JHEP {\bf 1301} (2013) 184
  doi:10.1007/JHEP01(2013)184
  [arXiv:1208.3674 [hep-th]].

\bibitem{ScaleVConformal4}
  A.~Dymarsky, Z.~Komargodski, A.~Schwimmer and S.~Theisen,
  ``On Scale and Conformal Invariance in Four Dimensions,''
  JHEP {\bf 1510} (2015) 171
  doi:10.1007/JHEP10(2015)171
  [arXiv:1309.2921 [hep-th]].
  
  \bibitem{Adler1}
  S.~L.~Adler,
  ``Consistency conditions on the strong interactions implied by a partially conserved axial vector current,''
  Phys.\ Rev.\  {\bf 137} (1965) B1022.
  doi:10.1103/PhysRev.137.B1022

\bibitem{Adler2}
  I.~Low,
  ``Adler’s zero and effective Lagrangians for nonlinearly realized symmetry,''
  Phys.\ Rev.\ D {\bf 91} (2015) no.10,  105017
  doi:10.1103/PhysRevD.91.105017
  [arXiv:1412.2145 [hep-th]].

\bibitem{ScalarSoft1}
  C.~Cheung, K.~Kampf, J.~Novotny and J.~Trnka,
  ``Effective Field Theories from Soft Limits of Scattering Amplitudes,''
  Phys.\ Rev.\ Lett.\  {\bf 114} (2015) no.22,  221602
  doi:10.1103/PhysRevLett.114.221602
  [arXiv:1412.4095 [hep-th]].

\bibitem{ScalarSoft2}
  C.~Cheung, K.~Kampf, J.~Novotny, C.~H.~Shen and J.~Trnka,
  ``A Periodic Table of Effective Field Theories,''
  JHEP {\bf 1702} (2017) 020
  doi:10.1007/JHEP02(2017)020
  [arXiv:1611.03137 [hep-th]].

\bibitem{ScalarSoft3}
  A.~Padilla, D.~Stefanyszyn and T.~Wilson,
  ``Probing Scalar Effective Field Theories with the Soft Limits of Scattering Amplitudes,''
  JHEP {\bf 1704} (2017) 015
  doi:10.1007/JHEP04(2017)015
  [arXiv:1612.04283 [hep-th]].

\bibitem{ScalarLie1}
  M.~P.~Bogers and T.~Brauner,
  ``Lie-algebraic classification of effective theories with enhanced soft limits,''
  JHEP {\bf 1805} (2018) 076
  doi:10.1007/JHEP05(2018)076
  [arXiv:1803.05359 [hep-th]].

\bibitem{AlgebraicClassification}
  D.~Roest, D.~Stefanyszyn and P.~Werkman,
  arXiv:1903.08222 [hep-th].
  
\bibitem{Son}
  D.~T.~Son,
  ``Low-energy quantum effective action for relativistic superfluids,''
  hep-ph/0204199.
  


\bibitem{Superfluid1}
  M.~Greiter, F.~Wilczek and E.~Witten,
  ``Hydrodynamic Relations in Superconductivity,''
  Mod.\ Phys.\ Lett.\ B {\bf 3} (1989) 903.
  doi:10.1142/S0217984989001400


\bibitem{FiniteTemperature}
  A.~Nicolis,
  arXiv:1108.2513 [hep-th].

\bibitem{Coset1}
  S.~R.~Coleman, J.~Wess and B.~Zumino,
  ``Structure of phenomenological Lagrangians. 1.,''
  Phys.\ Rev.\  {\bf 177} (1969) 2239.

\bibitem{Coset2}
  C.~G.~Callan, Jr., S.~R.~Coleman, J.~Wess and B.~Zumino,
  ``Structure of phenomenological Lagrangians. 2.,''
  Phys.\ Rev.\  {\bf 177} (1969) 2247.


\bibitem{Coset3}
  D.~V.~Volkov,
  ``Phenomenological Lagrangians,''
  Fiz.\ Elem.\ Chast.\ Atom.\ Yadra {\bf 4} (1973) 3.

  
    \bibitem{Zoology}
  A.~Nicolis, R.~Penco, F.~Piazza and R.~Rattazzi,
  ``Zoology of condensed matter: Framids, ordinary stuff, extra-ordinary stuff,''
  JHEP {\bf 1506} (2015) 155
  doi:10.1007/JHEP06(2015)155
  [arXiv:1501.03845 [hep-th]].



\bibitem{SuperfluidCoset}
  A.~Nicolis, R.~Penco and R.~A.~Rosen,
  ``Relativistic Fluids, Superfluids, Solids and Supersolids from a Coset Construction,''
  Phys.\ Rev.\ D {\bf 89} (2014) no.4,  045002
  doi:10.1103/PhysRevD.89.045002
  [arXiv:1307.0517 [hep-th]].
  

\bibitem{InverseHiggs}
  E.~A.~Ivanov and V.~I.~Ogievetsky,
  ``The Inverse Higgs Phenomenon in Nonlinear Realizations,''
  Teor.\ Mat.\ Fiz.\  {\bf 25} (1975) 164.

\bibitem{LowGoldstone}
  I.~Low and A.~V.~Manohar,
  ``Spontaneously broken space-time symmetries and Goldstone's theorem,''
  Phys.\ Rev.\ Lett.\  {\bf 88} (2002) 101602
  doi:10.1103/PhysRevLett.88.101602
  [hep-th/0110285].


\bibitem{Olver}
P.J. Olver, ``Applications of Lie Groups to Differential Equations", \textit{Springer} (1986)

\bibitem{Extended}
  K.~Hinterbichler and A.~Joyce,
  ``Goldstones with Extended Shift Symmetries,''
  Int.\ J.\ Mod.\ Phys.\ D {\bf 23} (2014) no.13,  1443001
  doi:10.1142/S0218271814430019
  [arXiv:1404.4047 [hep-th]].

\bibitem{VectorLie1}
  R.~Klein, E.~Malek, D.~Roest and D.~Stefanyszyn,
  ``A No-go Theorem for a Gauge Vector as a Space-time Goldstone,''
  arXiv:1806.06862 [hep-th].



\bibitem{Tasinato}
  J.~Chagoya and G.~Tasinato,
  ``A geometrical approach to degenerate scalar-tensor theories,''
  JHEP {\bf 1702} (2017) 113
  doi:10.1007/JHEP02(2017)113
  [arXiv:1610.07980 [hep-th]].
  
\bibitem{Galileon}
  A.~Nicolis, R.~Rattazzi and E.~Trincherini,
  ``The Galileon as a local modification of gravity,''
  Phys.\ Rev.\ D {\bf 79} (2009) 064036
  doi:10.1103/PhysRevD.79.064036
  [arXiv:0811.2197 [hep-th]].
  
\bibitem{Goon}
  G.~Goon, K.~Hinterbichler, A.~Joyce and M.~Trodden,
  JHEP {\bf 1206} (2012) 004
  doi:10.1007/JHEP06(2012)004
  [arXiv:1203.3191 [hep-th]].

\bibitem{Reunited}
  C.~de Rham and A.~J.~Tolley,
  ``DBI and the Galileon reunited,''
  JCAP {\bf 1005} (2010) 015
  doi:10.1088/1475-7516/2010/05/015
  [arXiv:1003.5917 [hep-th]].

\bibitem{DBIInflation1}
  E.~Silverstein and D.~Tong,
  ``Scalar speed limits and cosmology: Acceleration from D-cceleration,''
  Phys.\ Rev.\ D {\bf 70} (2004) 103505
  doi:10.1103/PhysRevD.70.103505
  [hep-th/0310221].

\bibitem{DBIInflation2}
  P.~Creminelli, R.~Emami, M.~Simonović and G.~Trevisan,
  ``ISO(4,1) Symmetry in the EFT of Inflation,''
  JCAP {\bf 1307} (2013) 037
  doi:10.1088/1475-7516/2013/07/037
  [arXiv:1304.4238 [hep-th]].

\bibitem{ConformalSuperfluid1}
  S.~Hellerman, D.~Orlando, S.~Reffert and M.~Watanabe,
  ``On the CFT Operator Spectrum at Large Global Charge,''
  JHEP {\bf 1512} (2015) 071
  doi:10.1007/JHEP12(2015)071
  [arXiv:1505.01537 [hep-th]].

\bibitem{ConformalSuperfluid2}
  A.~Monin, D.~Pirtskhalava, R.~Rattazzi and F.~K.~Seibold,
  ``Semiclassics, Goldstone Bosons and CFT data,''
  JHEP {\bf 1706} (2017) 011
  doi:10.1007/JHEP06(2017)011
  [arXiv:1611.02912 [hep-th]].

\bibitem{ConformalSuperfluid3}
  G.~Cuomo, A.~de la Fuente, A.~Monin, D.~Pirtskhalava and R.~Rattazzi,
  ``Rotating superfluids and spinning charged operators in conformal field theory,''
  Phys.\ Rev.\ D {\bf 97} (2018) no.4,  045012
  doi:10.1103/PhysRevD.97.045012
  [arXiv:1711.02108 [hep-th]].

\bibitem{ConformalSuperfluid4}
  A.~Esposito, S.~Garcia-Saenz and R.~Penco,
  ``First sound in holographic superfluids at zero temperature,''
  JHEP {\bf 1612} (2016) 136
  doi:10.1007/JHEP12(2016)136
  [arXiv:1606.03104 [hep-th]].
  
\bibitem{Sundrum}
  R.~Sundrum,
  hep-th/0312212.
  
\bibitem{CFT}
  A.~A.~Belavin, A.~M.~Polyakov and A.~B.~Zamolodchikov,
  Nucl.\ Phys.\ B {\bf 241} (1984) 333.
  doi:10.1016/0550-3213(84)90052-X
  
\bibitem{Blumenhagen:2009zz}
  R.~Blumenhagen and E.~Plauschinn,
  Lect.\ Notes Phys.\  {\bf 779} (2009) 1.
  doi:10.1007/978-3-642-00450-6
  
\bibitem{Finelli:2018upr}
  B.~Finelli, G.~Goon, E.~Pajer and L.~Santoni,
  ``The Effective Theory of Shift-Symmetric Cosmologies,''
  JCAP {\bf 1805} (2018) no.05,  060
  doi:10.1088/1475-7516/2018/05/060
  [arXiv:1802.01580 [hep-th]].


\bibitem{Nicolis:2011pv}
  A.~Nicolis and F.~Piazza,
  ``Spontaneous Symmetry Probing,''
  JHEP {\bf 1206} (2012) 025
  doi:10.1007/JHEP06(2012)025
  [arXiv:1112.5174 [hep-th]].

 \bibitem{BFSymmetry}
 Bernardo Finelli, unpublished.

\bibitem{Coleman:1970je}
  S.~R.~Coleman and R.~Jackiw,
  ``Why Dilatation Generators Do Not Generate Dilatations?,''
  Annals Phys.\  {\bf 67} (1971) 552.
  doi:10.1016/0003-4916(71)90153-9


\bibitem{Callan:1970ze}
  C.~G.~Callan, Jr., S.~R.~Coleman and R.~Jackiw,
  ``A New Improved Energy - Momentum Tensor,''
  Annals Phys.\  {\bf 59} (1970) 42.
  doi:10.1016/0003-4916(70)90394-5
  
\bibitem{Imaginary}
  S.~Garcia-Saenz and S.~Renaux-Petel,
  JCAP {\bf 1811} (2018) no.11,  005
  doi:10.1088/1475-7516/2018/11/005
  [arXiv:1805.12563 [hep-th]].
  
\bibitem{Signs}
  D.~Baumann, D.~Green, H.~Lee and R.~A.~Porto,
  Phys.\ Rev.\ D {\bf 93} (2016) no.2,  023523
  doi:10.1103/PhysRevD.93.023523
  [arXiv:1502.07304 [hep-th]].
  
\bibitem{EnergyAmplitude}
  A.~Nicolis, R.~Rattazzi and E.~Trincherini,
  JHEP {\bf 1005} (2010) 095
   Erratum: [JHEP {\bf 1111} (2011) 128]
  doi:10.1007/JHEP05(2010)095, 10.1007/JHEP11(2011)128
  [arXiv:0912.4258 [hep-th]].
  
\bibitem{GhostCondensate}
  N.~Arkani-Hamed, H.~C.~Cheng, M.~A.~Luty and S.~Mukohyama,
  ``Ghost condensation and a consistent infrared modification of gravity,''
  JHEP {\bf 0405} (2004) 074
  doi:10.1088/1126-6708/2004/05/074
  [hep-th/0312099].
  
\bibitem{Fermi}
  J.~Polchinski,
  In *Boulder 1992, Proceedings, Recent directions in particle theory* 235-274, and Calif. Univ. Santa Barbara - NSF-ITP-92-132 (92,rec.Nov.) 39 p. (220633) Texas Univ. Austin - UTTG-92-20 (92,rec.Nov.) 39 p
  [hep-th/9210046].


\bibitem{Vainshtein}
  N.~Kaloper, A.~Padilla, P.~Saffin and D.~Stefanyszyn,
  ``Unitarity and the Vainshtein Mechanism,''
  Phys.\ Rev.\ D {\bf 91} (2015) no.4,  045017
  doi:10.1103/PhysRevD.91.045017
  [arXiv:1409.3243 [hep-th]].

\bibitem{HigherOrderIH}
  R.~Klein, D.~Roest and D.~Stefanyszyn,
  ``Spontaneously Broken Spacetime Symmetries and the Role of Inessential Goldstones,''
  JHEP {\bf 1710} (2017) 051
  doi:10.1007/JHEP10(2017)051
  [arXiv:1709.03525 [hep-th]].
  
\bibitem{SuperfluidWess}
  L.~V.~Delacrétaz, A.~Nicolis, R.~Penco and R.~A.~Rosen,
  ``Wess-Zumino Terms for Relativistic Fluids, Superfluids, Solids, and Supersolids,''
  Phys.\ Rev.\ Lett.\  {\bf 114} (2015) no.9,  091601
  doi:10.1103/PhysRevLett.114.091601
  [arXiv:1403.6509 [hep-th]].
  
\bibitem{Bordin:2018pca} 
  L.~Bordin, P.~Creminelli, A.~Khmelnitsky and L.~Senatore,
  ``Light Particles with Spin in Inflation,''
  JCAP {\bf 1810}, no. 10, 013 (2018)
  doi:10.1088/1475-7516/2018/10/013
  [arXiv:1806.10587 [hep-th]].

\bibitem{Bruneton:2007si} 
  J.~P.~Bruneton and G.~Esposito-Farese,
  ``Field-theoretical formulations of MOND-like gravity,''
  Phys.\ Rev.\ D {\bf 76}, 124012 (2007)
  Erratum: [Phys.\ Rev.\ D {\bf 76}, 129902 (2007)]
  doi:10.1103/PhysRevD.76.129902, 10.1103/PhysRevD.76.124012
  [arXiv:0705.4043 [gr-qc]].
  
\bibitem{Berezhiani:2015bqa}
  L.~Berezhiani and J.~Khoury,
  ``Theory of dark matter superfluidity,''
  Phys.\ Rev.\ D {\bf 92} (2015) 103510
  doi:10.1103/PhysRevD.92.103510
  [arXiv:1507.01019 [astro-ph.CO]].
  
\bibitem{Khoury:2018vdv}
  J.~Khoury, J.~Sakstein and A.~R.~Solomon,
  ``Superfluids and the Cosmological Constant Problem,''
  JCAP {\bf 1808} (2018) no.08,  024
  doi:10.1088/1475-7516/2018/08/024
  [arXiv:1805.05937 [hep-th]].

\bibitem{Ferreira:2018wup} 
  E.~Ferreira, G.M., G.~Franzmann, J.~Khoury and R.~Brandenberger,
  arXiv:1810.09474 [astro-ph.CO].
    

  






\end{thebibliography}
\end{document}